\documentclass[letterpaper, 11 pt, onecolumn, draftclsnofoot]{IEEEtran}

\usepackage{amsthm}

\usepackage{upgreek}
\usepackage{stmaryrd}

\usepackage[usenames, dvipsnames]{color}

\usepackage{hyperref}
\hypersetup{hidelinks}
\usepackage{optidef}
\usepackage{graphics} % for pdf, bitmapped graphics files
\graphicspath{{figures/}}
\usepackage{amsmath} % assumes amsmath package installed
\usepackage{amssymb}  % assumes amsmath package installed
\usepackage{balance}
 
\usepackage{verbatim}

\usepackage{bm}

\makeatletter
\newcommand\gobblepars{%
    \@ifnextchar\par%
 {\expandafter\gobblepars\@gobble}%
{}}
\makeatother

\def\wham#1{\medbreak\pagebreak[3]%
\noindent\textbf{#1}\ \ \gobblepars}

\newcounter{rmnum}

\def\Ebox#1#2{%
\begin{center} 
\includegraphics[width= #1\hsize]{#2}\end{center}}

\newlength{\noteWidth}
\long\def\notes#1{\ifinner
             {\tiny #1}
             \else
             \marginpar{\parbox[t]{\noteWidth}{\raggedright\tiny #1}}
             \fi}
             
           \def\notes#1{\typeout{See notes!}}
\def\archive#1{}

\def\MC{\text{\MC}}

\newcommand{\field}[1]{\mathbb{#1}}

\def\Re{\field{R}}

\def\eqdef{\mathbin{:=}}

\usepackage{cleveref}
\Crefname{corollary}{Corollary}{Corollaries}
\Crefname{eqnarray}{eq.}{eqs.}
\Crefname{equation}{eq.}{eqs.}

\Crefname{figure}{Fig.}{Figs.}
\Crefname{tabular}{Tab.}{Tabs.}
\Crefname{table}{Tab.}{Tabs.}
\Crefname{lemma}{Lemma}{Lemmas}

\Crefname{theorem}{Thm.}{Thms.}
\Crefname{definition}{Definition}{Definitions}
\Crefname{section}{Section}{Sections}
\Crefname{proposition}{Prop.}{Propositions}
\Crefname{assumption}{Assumption}{Assumptions}
\Crefname{example}{Example}{Examples}

\def\barell{{\overline {\ell}}}

\def\bfmath#1{{\mathchoice{\mbox{\bfmath$#1$}}%
{\mbox{\boldmath$#1$}}%
{\mbox{\boldmath$\scriptstyle#1$}}%
{\mbox{\boldmath$\scriptscriptstyle#1$}}}}

\def\cX{c_{\text{\tiny X}}}
\def\cG{c_{\text{g}}}

\def\cdG{c_{{\text{\lower1pt\hbox{r}}}} }

\def\Toff{T_{\text{\it off}}}
\def\Ton{T_{\text{\it on}}}

 \def\FRAC#1#2#3{\genfrac{}{}{}{#1}{#2}{#3}}

\def\half{{\mathchoice{\FRAC{1}{1}{2}}%
{\FRAC{1}{1}{2}}%
{\FRAC{3}{1}{2}}%
{\FRAC{3}{1}{2}}}}

\def\clP{{\cal P}}

\def\clT{{T}}

\def\cX{c_{\text{\tiny X}}}
\def\cG{c_{\text{g}}}
\def\cdG{c_{{\text{\lower1pt\hbox{d}}}} }

\usepackage{bbold} 

\RenewEnviron{BaseMiniExclam}[7]{%
	\selectConstraintMult{#1}%
	\renewcommand{\localOptimalVariable}{#2}%
	\begin{subequations}
		\ifthenelse{\equal{#7}{b}}{\allowdisplaybreaks}%
		#4
		\begin{alignat}{5}
		\bodyobj{#2}{#3}{#6}{#5}
		\BODY
		\end{alignat}
	\end{subequations}%
	\setStandardMini
}

\makeatletter
\newcommand{\oset}[3][0ex]{%
  \mathrel{\mathop{#3}\limits^{
    \vbox to#1{\kern-2\ex@
    \hbox{$\scriptstyle#2$}\vss}}}}
\makeatother

 \def\MPCt{\tau}
 \def\MPCshift{t_s}
 \def\INITt{t_0}

\makeatletter
\def\thanks#1{\protected@xdef\@thanks{\@thanks
		\protect\footnotetext{#1}}}
\makeatother
\title{Model Predictive Control for Joint Ramping and Regulation-Type Service from Distributed Energy Resource Aggregations}
\author{\IEEEauthorblockN{\IEEEauthorblockA{\text{Joel~Mathias}, 
\text{Rajasekhar~Angluri}, \text{Oliver~Kosut}, 
 \text{and~Lalitha~Sankar}}
\text{School of Electrical, Computer, and Energy
			Engineering, Arizona State University, Tempe, AZ 85281, USA} \\
\{joel.mathias, rajasekhar.anguluri, lsankar, okosut\}@asu.edu}}
\begin{document}

\maketitle

%\tableofcontents

\begin{abstract}

 Distributed energy resources (DERs) such as grid-responsive loads and batteries can be harnessed to provide ramping and regulation services across the grid. This paper concerns the problem of optimal allocation of different classes of DERs, where each class is an aggregation of similar DERs, to balance net-demand forecasts. 
 The resulting resource allocation problem is solved using model-predictive control (MPC) that utilizes a rolling sequence of finite time-horizon constrained optimizations. This is based on the concept that we have more accurate estimates of the load forecast in the short term, so each optimization in the rolling sequence of optimization problems uses more accurate short term load forecasts while ensuring satisfaction of %transient performance and 
 capacity and dynamical constraints. Simulations demonstrate that the MPC solution can indeed reduce the ramping required from bulk generation, while mitigating near-real time grid disturbances.

\end{abstract}

\section{Introduction}

Massive renewable generation introduces volatility and uncertainty in the grid, leading to sharp ramps and disturbances in net demand (demand minus renewable generation). Traditionally, these ramps and disturbances have been addressed using fossil-based generators such as coal and diesel, which are costly, inefficient, and not environment friendly.

 The grid operator can tackle these challenges by intelligently allocating distributed energy resources (DERs), such as solar photovoltaic, wind turbines, battery storage, and flexible loads. \emph{We propose a methodology that leverages aggregate DER models to optimally allocate DERs to meet the ramps and disturbances in net-demand forecasts (net-demand is demand minus renewable generation) using a model predictive control (MPC) framework}.  We do so while including the dynamical behavior and capacity constraints of the DER aggregation.

The grid operator and DER resource aggregators form the agents in our proposed approach. Aggregators are entities that engage with the consumers to aggregate DERs to deploy them based on the requirements of the grid operator (e.g., EnergyHub, Enel X, and utility companies). The role of the grid operator (e.g., ERCOT) is to coordinate DER aggregations and bulk generation to procure and dispatch the resources required to balance supply and demand. The formulation in this paper focuses on DER aggregations: the precise way in which the control is designed at an individual level to achieve tracking is beyond the scope of this paper (see \cite{matbusmey23} and the references therein for individual-level control).

The control architectures presently used in the grid focus primarily on short-term optimization \cite{dorfler2019distributed}. Traditionally, economic dispatch with the participation of large generation plants is performed as a static optimization problem \cite{woowolshe13}. However, the optimal dispatch of DER aggregations needs to account for the dynamics of these aggregations, thereby requiring an optimal control framework. In such scenarios, the solution to short-term optimization is unlikely to be optimal in the long term. The control actions in the grid also vary across different time scales and information architectures: for example, day-ahead dispatch relies on forecasts over the next day, while ancillary services rely on near-real time estimates. 

To address these challenges, we introduce an MPC framework that approximates the solution to a long-run optimal control problem by solving a sequence of finite dimensional optimization problems.  We require two key ingredients in our framework: (a) models for the dynamical behavior of DER aggregations; and (b) accurate short-term predictions for the net-demand forecast. MPC has strong guarantees for transient operations and ensures constraint satisfaction, while also accounting for long-run optimality \cite{rawlings2017model}.

\begin{figure*}[tb]
	\vspace{-0.1em}
	\centering
	\includegraphics[width=.999\hsize, keepaspectratio]{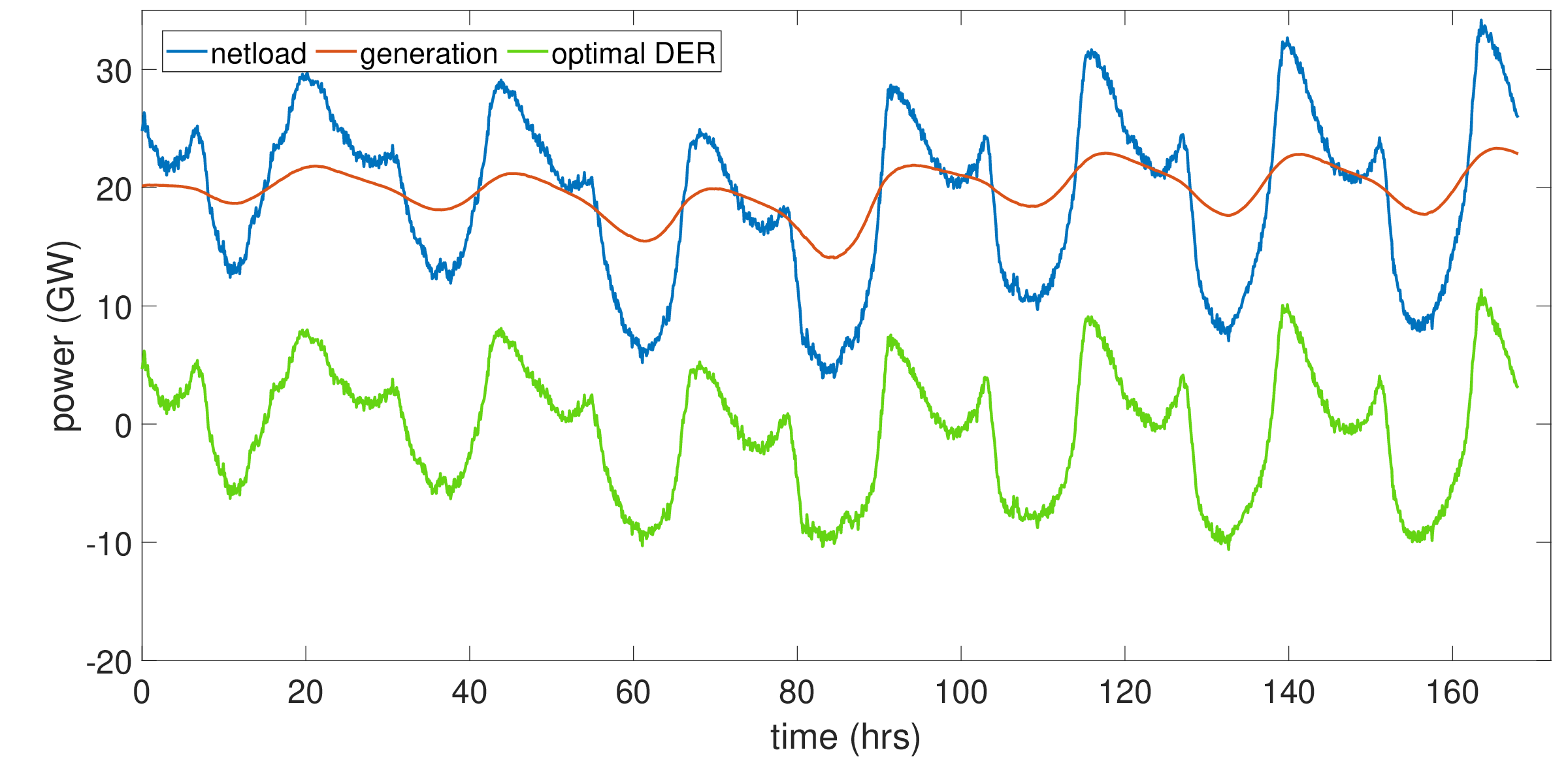}
	%\vspace{-.25em}
	\caption{Bulk generation, net-demand and the total power supplied by all classes of DER aggregations. The net-demand is from CAISO from September 1, 2023 through September 7 2023, with BPA's BRD disturbances (for the corresponding time period) coming in at 30 minute windows during each MPC update. The DER aggregations provide most of the ramping and regulation-type service required to balance the net load, thereby reducing the requirements from bulk generation.} % 
	\vspace{-.75em}
	\label{f:MPC_traj} 
\end{figure*}

\wham{Related Research}\label{sec: related research}
Finite-horizon convex optimization formulations for the control of DER aggregations are discussed in \cite{espalm20}: a centralized quadratic program is solved to generate different command signals for the different classes of DERs. The paper \cite{bencolmal19} solves a non-linear AC optimal power flow (OPF) with the participation of an aggregation of DERs, specifically thermostatically controlled loads: the problem is modeled as a Markov decision process (MDP) and is reformulated as a finite-horizon convex program. The cost formulations in this paper are related to \cite{cammatkiebusmey18, matmeybalans21}, which also consider finite horizon convex programs to allocate load aggregations.  However, all these approaches focus solely on short-term, finite horizon optimization. Model predictive control is used for economic dispatch in power networks in \cite{kohlerMPC}; however, the formulation does not include DERs.   

 Moreover, these prior works do not consider the near-real time disturbances entering the grid in their optimal control formulations. We consider a general constrained MPC framework for  allocation of DER aggregations: this optimization strategy addresses both transient grid disturbances and accounts for net-demand forecasts over a longer horizon, without violating capacity constraints of the DERs.  A key ingredient of our set-up is the use of convex costs on the state of charge (SoC) of the DER aggregation with hard constraints based on power and SoC capacity. By regularly updating the forecasts, we can use DERs to provide both ramping and regulation-type services to the grid. This reduces the requirement from bulk generation. The simulations in \Cref{s:numerics} illustrate the efficacy of the control design in jointly extracting dispatch and regulation-type services from the DERs.

\wham{Notation}  
\\
\noindent
$\MPCt$: control time horizon for an MPC iteration indexed by $t \in \{\INITt,t_0+1,\ldots,\INITt+\MPCt-1\}$.\\
$\INITt$: starting time for an MPC iteration.\\
$\MPCshift$: time shift for $t_0$ between successive MPC iterations.\\
$\ell(t)$ : forecast of net-demand at time $t$.  \\
%$\ell^\text{RT}(t)$: net-demand at time $t$ (measured at real-time). \\ 
$g(t)$ : power from bulk generation (without renewables).
\\
$M$: number of DER aggregations, indexed by $i \in \{1,\ldots,M\}$.
\\
$x_i (t)$ : SoC for DER aggregation $i$ (units: GWh). \\
$p_i(t)$ : power output of aggregator $i$ (units: GW).

\begin{figure*}[tb]
	\vspace{-0.1em}
	\centering
	\includegraphics[width=.999\hsize, keepaspectratio]{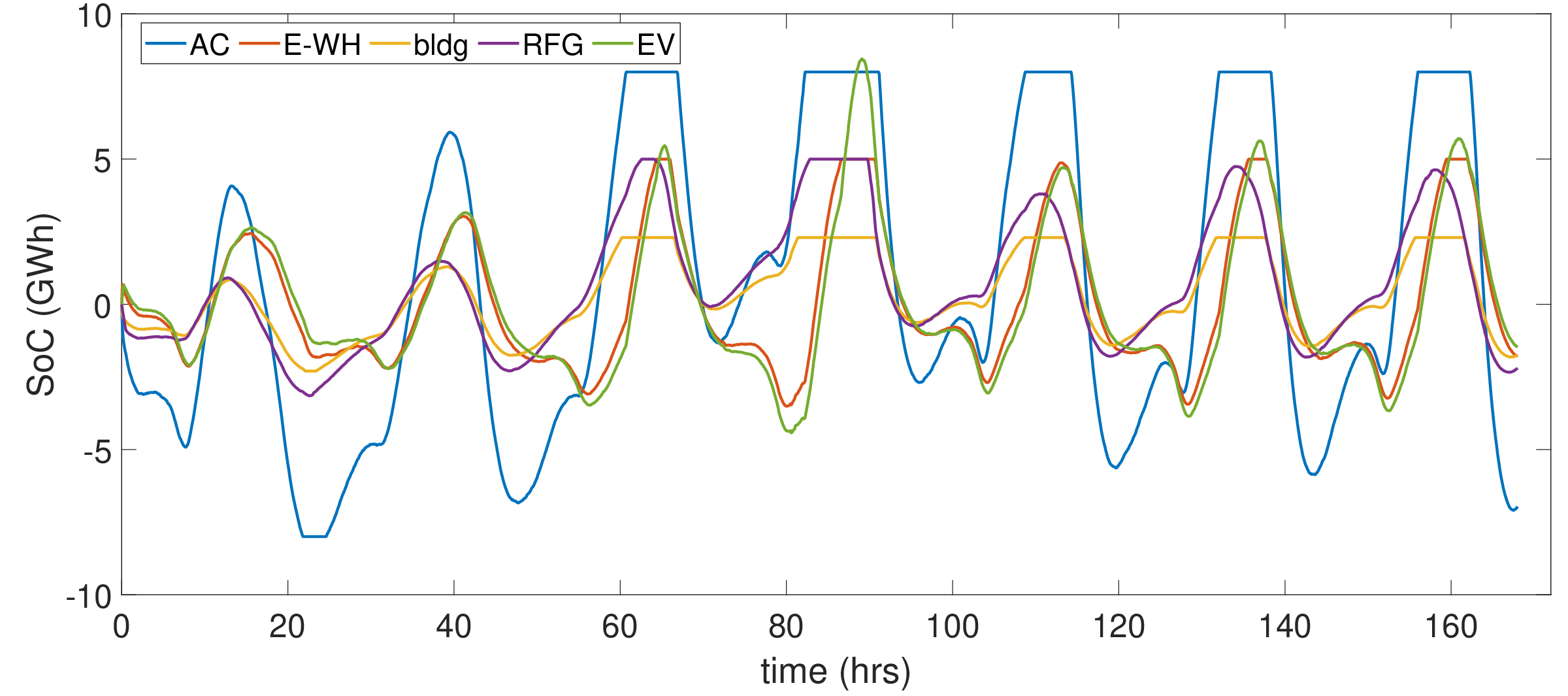}
	%\vspace{-.25em}
	\caption{SoC trajectories via MPC from five classes of DER aggregations that are deployed to meet net-demand forecast for CAISO from September 1 through September 7, 2023. Each optimization problem has a horizon $\MPCt = 24$ hours. The time shift between sucessive optimization problems is $\MPCshift= 30$ minutes. BPA's BRD from September 1 through September 7, 2023, are added to the net-demand forecast every 30 minutes: these comprise of ``near real-time" disturbances to the net-demand. The figure shows that some classes of DERs can charge to their maximum SoC capacity: particularly noticeable for ACs and bldgs.}
	\vspace{-.75em}
	\label{f:SoC} 
\end{figure*}

\wham{Organization}
The problem setup is explained in \Cref{s:prelim};  
in particular, the linear dynamical models for the DER aggregations are introduced in \Cref{s:ves_models}, while the cost functions associated with the optimal control framework are introduced in  \Cref{s:sdd}. 
The model predictive control architecture is introduced in \Cref{s:mpc}.  \Cref{s:numerics} provides results from simulation experiments, illustrating the application of the control design in realistic settings.  Conclusions and directions for future research are provided in \Cref{s:con}.

\section{Problem Setup}
\label{s:prelim}

We now introduce models for the control architecture, aggregate DER dynamics, and the optimization objective. 

\subsection{Generalized linear battery models}
\label{s:ves_models}
We consider linear models for aggregate DER dynamics, which are well studied literature (see \cite{haosanpoovin15, hugdompoo16, ma2011decentralized}). 
 Following \cite{haosanpoovin15}, we adopt the generalized battery models for DER aggregations with known constraints on the power and the state of charge. The SoC of the $i$-th DER aggregagation, $i \in \{1,\ldots,M\}$, obeys:
\begin{subequations}
\label{e:DER_model}
\begin{align}
\label{e:SoC_ODE}
x_i(t+1)  & = \alpha_i x_i(t)   - \beta p_i(t),\\ 
\label{e:SoC_cap}
|x_i(t)| & \le C_i, \\
\label{e:Power_cap}
-\eta_i^- & \le p_i(t) \le \eta_i^+, 
\end{align}
\end{subequations}
where the state $x_i(t)\in \Re$ models state of charge, and $\alpha_i\geq 0$ is a leakage parameter. For example, for aggregations of loads such as air conditioners and water heaters, $\alpha_i$ corresponds to the thermal time constant. The nominal SoC is defined to be 0 (nominal refers the behavior of the DER aggregation when the DERs are not participating in dispatch or regulation). The (virtual) power supplied at time $t$ by the $i$-th DER aggregation is $p_i(t) \in \Re$: this is the nominal power consumption of the DER aggregation minus the total power consumed by the aggregation. The constant $\beta$ is in units of time.

The appendix provides details on the derivation of the capacity limits for an aggregation of grid-responsive loads.

\subsection{DER allocation: an optimal control approach}
\label{s:sdd}

We formulate the resource allocation problem as an infinite-horizon optimal control problem with the objective:  
\begin{equation}
\lim_{\clT\to\infty} \sum_{t=0}^{\clT}    \big[   \cG (g(t) ) + \cX(x(t)) \bigr] \,,
    \label{e:obj}
\end{equation}
where the state cost $\cX$ is given by 
\begin{equation}
\cX(x) \eqdef  \sum_{i=1}^{M}   c_i(x_i) = \sum_{i=1}^M \half \kappa_i x_i^2,
\label{e:cX}
\end{equation}
and the generation cost $c_g$ is given by 
\begin{align}
\begin{split}
c_g(x) &= \half \kappa_g (x - \barell)^2, %\\
%c_d(x,y) &= \half \kappa_d (x-y)^2   ,\qquad x,y \in \Re,
\end{split}
\label{e:genCost_disc}
\end{align}
with $\barell\eqdef \clT^{-1}\sum_0^{\clT-1}\ell(t)$, in which $\ell(t)$ denotes the net-demand forecast. $\kappa_i\geq 0$ and $\kappa_g \geq 0$ are design parameters.

The net-demand forecast $\ell(t)$ satisfies 
\begin{equation}
\ell(t)=g(t)+p_\sigma (t)  \,,\qquad 0\le t\le \clT
    \label{e:supply=demand}
\end{equation}
where $p_\sigma (t) = \sum_i p_i(t)$ is the total power supplied by all the DERs, and $g(t)$ is the power generated by traditional resources. The initial condition $x_i(0)$ is known for each class. 

\begin{figure*}[tb]
	\vspace{-0.1em}
	\centering
	\includegraphics[width=.999\hsize, keepaspectratio]{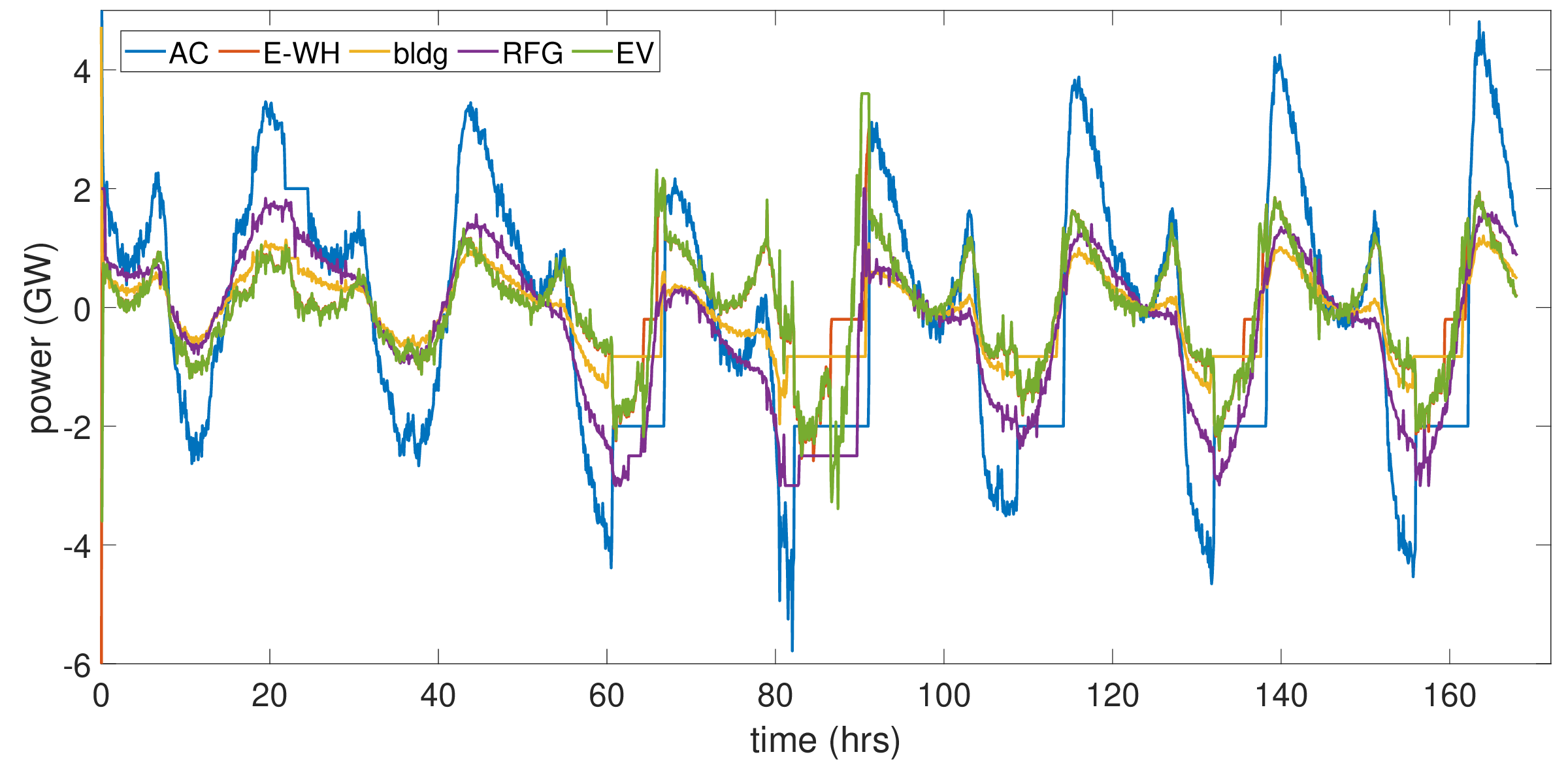}
	%\vspace{-.25em}
	\caption{Power trajectories for the five classes of DER aggregations obtained via the MPC solution. The DER aggregations provide both ramping and regulation-type service to the grid. The faster fluctuations are due the need to cancel the near-real time (forecast) disturbances because of the BPA BRD signal.}  
	\vspace{-.75em}
	\label{f:Power} 
\end{figure*}

\section{Model Predictive Control Architecture}
\label{s:mpc}

Solving the infinite horizon optimization problem described by \eqref{e:obj} in the previous section is not feasible for the following reasons: (i) we may not have accurate long-term estimates of DER models (e.g., $\alpha_i$) and constraints (e.g., $C_i$) and net-demand forecasts ($\ell(t)$), but have accurate estimates in the short term; and (ii) the capacity-constrained infinite-horizon optimization problem is not computationally tractable. Model predictive control is a well-established control architecture in this setting to convert the infinite-horizon problem to a finite-horizon problem with regular updated forecast  $\ell(t)$ \cite{rawlings2017model}.

Model predictive control is an iterative technique that requires a   look-ahead time horizon $\MPCt$ and a time shift $\MPCshift$. %, and ``$\MPCt$-terminal cost'' $c^\bullet$.  
For an MPC iteration with starting time $\INITt\ge 0$,  the input $p(\INITt)$ is obtained through the following methodology. \\ Step (i): First, given a starting state $x(\INITt) \in \Re ^M$,   we solve the following convex optimization problem for the finite time horizon $[\INITt,\INITt+\MPCt]$:
\vspace{-0.1in}
\begin{mini!} 
	{g, x, p}{ \begin{aligned} \begin{split}  
	\sum_{t=\INITt}^{\INITt+\MPCt-1}    \big[   \cG (g(t) ) &+ \cX(x(t)) \big] \\ \end{split} \end{aligned} }{}{}
	{\label{qp19}}{}
	\addConstraint{\ell(t)}{=g(t)+p_\sigma (t)}
	{\label{e:balancecons}}{}
	\addConstraint{ {x}_i(t+1)}{= \alpha_i x_i(t) - \beta p_i(t)}
	{\label{e:soccons}}{}
%	\addConstraint{\text{All constraints on $\bfmx_i, \bfmz_i,  \clL_i$}}
	\addConstraint{-C_i}{\leq x_i(t)\leq C_i}
        {\label{e:SoCcap}}{}
	\addConstraint{-\eta^-_i}{\leq p_i(t)\leq \eta^+_i}.
        {\label{e:PowerCap}}{}
%	\addConstraint{\dot\clL_i(t)}{=- \alpha_i \clL_i(t) - z_i(t)}
%	\addConstraint{b_{i-}}{\leq \clL_i(t)} {\leq b_{i+}}
%	\addConstraint{-\eta_{i-}}{\leq z_i(t)} {\leq \eta_{i+}}
	%\addConstraint{0}{=\int_{t=0}^{\clT}z_i(\tau)d\tau}{} 
\end{mini!}

Step (ii): The optimal power (input) for an MPC iteration starting at $\INITt$ is the solution $p_i(t)$ for $t \in \{\INITt, \INITt + 1, \ldots, \INITt + \MPCshift\}$ and $i \in \{1,\ldots,M\}$. 

Step (iii): Successive MPC iterations are solved by periodically updating the starting time $\INITt$ at intervals of $\MPCshift$, with $0 < \MPCshift \le \MPCt$. That is, for an MPC iteration with starting time $\INITt$, the starting time for the next MPC iteration is $\INITt + \MPCshift$.

\section{Simulations}
\label{s:numerics}

The simulations detailed here include a single class of bulk generation and five classes of DERs, each aggregating millions of DERs. The DER classes are as follows: air conditioners (ACs), electric water heaters (E-WHs), building HVACs (bldgs), refrigerators (RFGs), and electric vehicles with 100 kWh lithium ion batteries (EVs). Tab.~I provides details about the type and number of DERs as well as the model parameters for each aggregation. The number of DERs in each aggregation are extrapolated based on the population of California. We compute $\alpha$ values and the capacity limits for the different DER aggregations using data from the following papers: for ACs, E-WHs, and RFGs \cite{matdyscal15}, for bldgs \cite{hugdompoo16}, and for EVs \cite{ma2011decentralized}. See appendix for the precise formulae.

The net load trajectory considered in our experiments is from publicly available  California Independent System Operator (CAISO) dataset for September 2023 \cite{CAISO}. Furthermore, we model additive (forecast) disturbances to this net load using the balancing reserves deployed (BRD) in September 2023 by the Bonneville Power Administration (BPA) \cite{BPA}. BRD is considered a type of regulation signal. Both these datasets have a 5-minute resolution.

\begin{table}[b]
\vspace{-0.75 em}
	\centering
	\label[table]{tab:RA}
	\begin{tabular}{|| l c c c c c r ||}
		Par. & Unit & DER1 & DER2 & DER3 & DER4 & DER5 \\
		Type & --- & ACs & E-WHs & bldgs & RFGs & EVs \\
		N & million & 10 & 10 & 1 & 10 & 1   \\
		$\alpha_i$ & ---  & 0.98 & 0.99 & 0.97 & 0.96 & 0.99  \\
		$C_i$ & GWh &  8 & 5 & 2.3 & 5 & 50  \\
            $\eta^+_i$ & GW & 20 & 4 & 103 & 2 & 3.6 \\
            $\eta^-_i$ & GW & 30 & 50 & 3 & 3 & 3.6\\
            $\kappa_i$ & --- & 1 & 2 & 5 & 5 & 2 \\
            $\beta$ & s & 300 & 300 & 300 & 300 & 300
	\end{tabular}
	\bigskip
\caption{Parameters for each class of DER aggregation, corresponding to the discrete-time linear dynamical model in equation \eqref{e:DER_model}.}
\end{table}

\Cref{f:MPC_traj} shows the net load, the output from traditional generation, and the cumulative power supplied by the DERs over a seven day duration following MPC (with net-demand data corresponding to September 1 through September 7, 2023). The finite time horizon for each MPC iteration ($\MPCt$) is 24 hours, while the time shift between the starting time of successive MPC iterations ($\MPCshift$) is 30 minutes. This implies that forecast disturbances (taken from BPA's BRD signal) are added to the CAISO's day-ahead net-demand forecast every 30 minutes (in 30 minute rolling windows). Observe that the bulk generation is smooth without any sharp ramps. The total power supplied by the DER aggregations provides most of the balancing required to smooth the ramps as well as the additive disturbances (from BPA's BRD) in the net-demand trajectory.

The SoC trajectories of the different DER aggregations deployed via MPC are shown in \Cref{f:SoC}, while the corresponding power trajectories are shown in \Cref{f:Power}. The DERs charge during times of low net demand (i.e., they consume more energy with respect to nominal), so that they can supply energy (consume less than nominal) when the net demand ramps up. The fact that the DER aggregations are providing regulation-type services can be seen in the fast fluctuations of the power trajectories in \Cref{f:Power}. The capacity constraints are satisfied for all the DER classes.

\section{Conclusions and Future Work}
\label{s:con}

We have shown the efficacy of MPC-based control design for the optimal allocation of DER aggregations to meet net-demand: the MPC design is ideal as it uses the more accurate net-demand forecasts available in the short-term for resource allocation, ensures constraint satisfaction, while still accounting for long-run optimality. The simulations demonstrate that the DER aggregations can mitigate the ramping and regulation-type service required from bulk generation.

Learning an accurate terminal penalty for the MPC problem based on statistical learning approaches will be explored in future work: a particular challenge is to incorporate accurate long-term models for net-demand. While MPC mitigates some of the uncertainties involved in estimates of the net-demand and model parameters (via periodic updates), a rigorous approach to robust control design remains a topic for future work.

\appendix
\label[appendix]{s:cap}

%This section consists of a review and extensions of the notions of capacity in terms of energy and power for virtual energy storage obtained from loads. 

Below we derive the formulae for the power and capacity limits for an aggregation of $N$ DERs. While the theory is elucidated using the specific example of an aggregation of heating loads (e.g., grid-responsive E-WHs), extensions to other classes of DER aggregations are straightforward. Each aggregation consists of similar types of DERs.

\wham{Individual load model} We assume the idealized settings of \cite{haosanpoovin15}, which ignore disturbances. The linearized discrete-time dynamics of the $i$-th (heating) load are given by:
\begin{equation}
\begin{aligned}
\Theta_i(t+1)   &= \lambda (\Theta_i(t) - \Theta_i^a) + \gamma \beta M_i(t) \clP_m. 
\end{aligned}
\label{e:tcltempM}
\end{equation}
with constants $(\lambda \in [0,1])$ proportional to the thermal time constant of the load, $\gamma$ being the thermal capacitance divided by the coefficient of performance, and $\beta$ the sampling time (see \cite{haosanpoovin15, matdyscal15} for details).
 $\Theta_i(t) $ is the internal temperature, $\Theta_i^a$ is  the ambient temperature, and $M_i(t)\in\{0,1\}$ is  the power mode of the load (``on'' is indicated by $M_i(t)=1$).   $\clP_m$ is the power consumed when $M_i(t)=1$.  
$\Theta(t)$ decreases when the power is off ($M_i(t) = 0$),
 and
increases when the power is on ($M_i(t) = 1$). The on-off behavior regulates the internal temperature ($\Theta_i(t)$) to be between the upper limit ($\Theta_+$) and the lower limit ($\Theta_-$). The set-point is 
$
\Theta_s \eqdef \frac{1}{2}(\Theta_+ - \Theta_-).
$

\iffalse
\begin{figure}[th]
	\Ebox{0.9}{energy_def}
	\vspace{-.5em}
	\caption{Hysteresis control in a thermostatically controlled load (in continuous time) and the resulting dynamics of internal temperature and power (taken from \cite{JoelMathiasThesis21}).}
	\label{f:energy_def} 
	\vspace{-.75em}
\end{figure}
\fi
%%%%
In a stationary, nominal setting, the off-duration of the load is denoted as $\Toff$ and the on-duration as $\Ton$.
\iffalse
are computed from equation (\ref{e:tcltempM}) as follows:
\begin{subequations}
\begin{align}
\lambda \Toff &= \ln \Bigl(1  + \frac {\Theta_+ - \Theta_-}{\Theta_- -\Theta_a}\Bigr), \\
\lambda \Ton &= \ln 
\left (
1  + \frac {\Theta_+ - \Theta_-}{   {\gamma \clP_m}/{\lambda}- (\Theta_+ -\Theta_a )}   %\frac{\gamma \clP_m}{\lambda}
\right ).
\end{align}
\end{subequations}
\fi
The average power over time is given by,
\[
\clP_0 = \clP_m \frac{\Ton}{\Ton + \Toff}.
\]

\wham{Power capacity}
%\label{s:powerCap} 
Here, we explain the notion of power capacity for a collection of homogeneous loads.   The definition of average power for the collection is meaningful in a stationary setting (no disturbances): for $N$ loads, the average power is $N\clP_0$; this is regarded as the nominal or baseline power consumption. 

 The upper and lower power limits are denoted  $\eta_+ $ and  $\eta_- $, which are the maximum power supplied and consumed by the aggregation (with respect to a baseline), respectively.  If we turn on all the loads, then this is analogous to a (virtual) battery charging at rate $\eta_-$, while turning off all the loads implies that the (virtual) battery is discharging at rate $\eta_+$.  Consequently, for a homogeneous aggregation, we have $\eta_+ = N \clP_0$ and $\eta_- = N (\clP_m - \clP_0)$.

\wham{SoC Capacity}
%\label{s:therm_energyCap} 
The SoC for a collection of $N$ homogeneous loads is the thermal energy stored in the collection, normalized such that the nominal value is zero. The relation between the SoC and the internal temperature is given as follows:
\begin{equation}
\label{e:SOC_def}
x(t) = \sum_{i=1}^N \frac{1}{\gamma} (\Theta_i(t) - \Theta_s)
\end{equation}
The SoC dynamics follow the first order linear dynamical system given in \eqref{e:SoC_ODE}, with the capacity bound $|x(t)| \le C$ for all $t$. Comparing \eqref{e:SoC_ODE} with \eqref{e:tcltempM} gives $\alpha = \lambda$.

For $N$ homogeneous loads, the limit $C$ is reached when the internal temperature of each load is at the maximum limit $\Theta_+$ (or conversely, each load is at the minimum, $\Theta_-$). Consequently, it follows from \eqref{e:SOC_def} that

\begin{equation}
C = \frac {N} {\gamma} (\Theta_+- \Theta_s) = \frac{N}{2 \gamma} (\Theta_+- \Theta_-)
\label{e:SoC_E0}
\end{equation}

\bibliographystyle{abbrv}
\bibliography{strings,markov,q,extras,PolicyCollapseExtras,CollapseExtras}  

\end{document}